\documentclass[preprint2]{aastex6}

\usepackage{amsmath,graphicx,bm,amsbsy,color,natbib,subfigure,enumerate}
\usepackage{lipsum}
\usepackage{epstopdf}
\pdfoutput=1

\newcommand{\C}{\mathbf{C}}
\newcommand{\beq}{\begin{equation}}
\newcommand{\eeq}{\end{equation}}

\newcommand{\Eye}{\mathbf{I}}
\newcommand{\x}{\mathbf{x}}

\newcommand{\N}{\mathbf{N}}
\newcommand{\n}{\mathbf{n}}

\newcommand{\base}{\mathbf{b}}

\newcommand{\A}{\mathbf{A}}

\newcommand{\PSF}{\mathbf{P}} 

\newcommand{\y}{\mathbf{y}}

\newcommand{\D}{\mathbf{D}}

\newcommand{\shat}{\hat{\mathbf{s}}}
\newcommand{\xhat}{\widehat{\x}}

\newcommand{\ANA}{\A^\dagger \N^{-1} \A}
\newcommand{\hiMpc}{\,$h$\,Mpc$^{-1}$}

\begin{document} 

\title{Redundant Array Configurations for 21\,cm Cosmology}

\author{Joshua~S.~Dillon\altaffilmark{1,2,3}}
\author{Aaron~R.~Parsons\altaffilmark{1,2}}
\email{jsdillon@berkeley.edu}

\altaffiltext{1}{Department of Astronomy, UC Berkeley, Berkeley, CA}
\altaffiltext{2}{Radio Astronomy Laboratory, UC Berkeley, Berkeley, CA}
\altaffiltext{3}{Berkeley Center for Cosmological Physics, Berkeley, CA}


\begin{abstract}
Realizing the potential of 21\,cm tomography to statistically probe the intergalactic medium before and during the Epoch of Reionization requires large telescopes and precise control of systematics.  Next-generation telescopes are now being designed and built to meet these challenges, drawing lessons from first-generation experiments that showed the benefits of densely packed, highly redundant arrays---in which the same mode on the sky is sampled by many antenna pairs---for achieving high sensitivity, precise calibration, and robust foreground mitigation. In this work, we focus on the Hydrogen Epoch of Reionization Array (HERA) as an interferometer with a dense, redundant core designed following these lessons to be optimized for 21\,cm cosmology. We show how modestly supplementing or modifying a compact design like HERA's can still deliver high sensitivity while enhancing strategies for calibration and foreground mitigation. In particular, we compare the imaging capability of several array configurations, both instantaneously (to address instrumental and ionospheric effects) and with rotation synthesis (for foreground removal). We also examine the effects that configuration has on calibratability using instantaneous redundancy. We find that improved imaging with sub-aperture sampling via ``off-grid'' antennas and increased angular resolution via far-flung ``outrigger'' antennas is possible with a redundantly calibratable array configuration.
\end{abstract}

\maketitle
  

\section{Introduction} \label{sec:intro}
The quest to detect and characterize the 21\,cm signal from neutral hydrogen during the Cosmic Dawn up through the Epoch of Reionization (EoR) has driven the design and construction of a generation of radio interferometers. The breadth of designs of these telescopes, which include the Low Frequency Array (LOFAR; \citealt{LOFAR2013}), the Giant Metrewave Radio Telescope (GMRT; \citealt{newGMRT}), the Donald C. Backer Precision Array for Probing the Epoch of Reionization (PAPER; \citealt{PAPER}), and the Murchison Widefield Array (MWA; \citealt{MWAdesign,TingaySummary,BowmanMWAScience}), reflects different strategies for detecting the signal and separating it from bright astrophysical foregrounds. Because those foregrounds---mostly synchrotron emission from our Galaxy and comparatively nearby galaxies---are four to five orders of magnitude brighter than the cosmological signal \citep{GSM}, their statistical separation is of paramount importance. Despite the difficulties, the scientific payoff from a successful observation will be considerable. The 21\,cm signal contains vast amounts of information about formation and evolution of the first stars, galaxies, and black holes \citep{FurlanettoReview, miguelreview, PritchardLoebReview, SaleemEoRChapter, aviBook} and could one day prove the most sensitive test of $\Lambda$CDM \citep{Matt3,Yi,ClesseBackgroundReionizationOmniscopes}.

As the next generation of 21\,cm observatories like the Hydrogen Epoch of Reionization Array (HERA; \citealt{PoberNextGen}; \citealt{HERAOverview}) and the low-frequency portion of the Square Kilometer Array (SKA-low; \citealt{LeonCosmicDawnEoRSKA}) move from design to construction, it is essential to reflect upon the lessons of the first generation. The strategies they have employed to separate the signal from the foregrounds rely on the intrinsic spectral smoothness of the foregrounds. We know, however, that spectrally smooth foregrounds can create spectrally complicated three dimensional maps due to the intrinsic chromaticity of inteferometric measurements \citep{Dattapowerspec,AaronDelay,VedanthamWedge,MoralesPSShapes,Hazelton2013,CathWedge,ThyagarajanWedge,EoRWindow1,EoRWindow2}. The effect, known as the ``wedge'' for the shape of foreground contamination in cylindrically binned power spectra \citep{PoberWedge}, makes that separation harder. One can attempt to subtract a model of foregrounds propagated through a model of one's instrument \citep{bernardi,FHD} and then subtract the remaining residuals statistically as the LOFAR team is attempting \citep{ChapmanGMCA,BonaldiCCA}, although this remains a considerable challenge.

One can also simply give up on modes within the maximal ``horizon wedge'' (corresponding to the delay of a point source at the horizon for a given baseline) and filter them out. This foreground avoidance strategy, championed by PAPER with its delay spectrum approach \citep{AaronDelay}, sacrifices sensitivity (especially on the long baselines which exhibit the most spectral structure; \citealt{PoberNextGen}) for robust foreground excision and has enabled the most sensitive power spectrum limits in \citet{PAPER32Limits}, \citet{DannyMultiRedshift}, and most recently in \citet{PAPER64Limits}. PAPER's highly redundant, compact configuration minimizes the cost of foregoing imaging by working only within the ``EoR window,'' the region putatively uncontaminated by the foreground wedge.

The MWA, in contrast, uses its moderately dense configuration and imaging capability to pursue a hybrid approach \citep{X13, EmpiricalCovariance, CHIPS}, first subtracting foregrounds from maps (or visibilities) before projecting out the most contaminated wedge modes. In this case, the goal of foreground subtraction is not an increase in sensitivity by recovering more foreground contaminated modes, but rather to minimize the leakage of foreground contamination out of the wedge and into the EoR window \citep{EmpiricalCovariance, PoberSidelobe}. Since the foregrounds are so bright, any small leakage can swamp the 21\,cm signal and lead to systematic errors that do not integrate down with additional observation. Subtracting foregrounds can both ameliorate those errors and get us closer to working within the wedge.

That said, a primary reason why spectrally smooth foregrounds can lead to contamination beyond the wedge is miscalibration \citep{ShawCoaxing}. Spectrally smooth foregrounds observed by an instrument with a subtly unsmooth spectral response can create foreground contamination inseparable from the cosmological signal. Traditionally, radio telescopes are calibrated by pointing at bright point sources with well-known spectra \citep{ThompsonMoranSwenson}. This is more complicated for instruments with the wide fields of view like LOFAR and especially the MWA, which must calibrate on a point source model with thousands of sources\footnote{Observed with a beam power pattern that may vary from time to time and antenna to antenna, necessitating simultaneous beam and antenna gain calibration.} and continually refine those models with new and deeper observations with better-understood maps \citep{selfcal,InitialLOFAR1}.

In an array optimized for instantaneous redundancy like PAPER, the fact that each mode on the sky is measured multiple times allows for the simultaneous solution of visibilities and calibration parameters at a singe frequency \citep{Wieringa,redundant,NonLinearRedCal}. All that remains is bandpass calibration of the entire array, rather than each element individually. The redundant calibration approach, by not requiring a sky model for more than a single overall bandpass, is a powerful alternative to traditional radio inteferometric calibration. It was implemented successfully by \citet{MITEoR} and proved a key improvement to PAPER analysis that enabled the \citet{PAPER64Limits} limits.

The desire to maximize sensitivity on the short baselines that have the least wedge contamination by sampling them redundantly is generally thought to be at odds with high fidelity imaging. It seems, on the face of it, that we must choose between the two strategies for calibration and foreground mitigation. This paper seeks to answer the question: can we have our cake and eat it too? Can we build a redundant array with high sensitivity on short baselines that enables both strategies for calibration and foreground mitigation? To answer that question in a specific case, we start with the core design for HERA---331 hexagonally packed 14\,m dishes (\citealt{HERADish1,HERADish2,HERADish3}; Patra et al.~\emph{in~prep.})---and examine how modest additions or rearrangements of antennas can improve mapmaking ability without sacrificing the precise calibratability or high sensitivity enabled by redundancy.

Previous array configuration studies have focused on other goals. \citet{UeLiPAST}, \citet{MWAdesign}, and \citet{WijnholdsCalibratability} look at array configurations designed to suppress sidelobes, but do not consider redundancy or optimizing EoR window sensitivity. \citet{FFTT2} explore the broad class of array designs that can be correlated with the fast Fourier transform (FFT) but did not pursue any particular optimization scheme. \citet{AaronSensitivity} explore the benefits of instantaneous redundancy, but do not optimize for redundant calibratability or imaging capability. More recently, \citet{GreigSKADesign} focus on varying levels of array compactness and their effects on sensitivity as manifested by errors on a parametrized model of reionization, but does not consider the benefits of redundancy for controlling systematics.

In this work, we assess the mapmaking capability and redundant calibratability of three variants of the HERA configuration, developing criteria for assessing array configurations that address not just sensitivity but also the all-important control of systematic effects. In Section \ref{sec:design}, we discuss the motivations for the design of the HERA core and present variants that increase its ability to make widefield, high-resolution maps without sacrificing sensitivity. In Section \ref{sec:mapmaking}, we look in detail at both the qualitative and quantitative effects that array configuration has on mapmaking. Next, in Section \ref{sec:calibration}, we examine the relative redundant calibratability of the arrays, including the distant ``outrigger'' antennas. Finally, in Section \ref{sec:summary} we conclude with a discussion of the lessons learned for HERA in particular and for the next generation of 21\,cm interferometers more generally.


\section{Array Design} \label{sec:design}

We begin our analysis of modified HERA configurations by presenting three variants to the HERA core design, a hexagonally packed core of 14\,m dishes (see Figure \ref{fig:arrays} and, for more detail, \citealt{HERAOverview}). The need to achieve a large collecting area cheaply while maximizing the sensitivity to short baselines drives this basic design, since the benefits of coherent averaging generally outweigh the additional sample variance due to a limited field of view \citep{AaronSensitivity,PoberNextGen}. In fact, its 21\,cm focused design allows it a similar performance to the SKA-low, especially when foregrounds must be avoided rather than subtracted  \citep{PoberNextGen, 21CMMC, AaronNextGenXRay}. HERA's zenith-pointing dishes, while restricting the survey area to a single stripe in declination, minimize the so-called ``pitchfork effect'' \citep{NithyaPitchfork,NithyaPitchforkConfirmation} in which diffuse emission near the horizon contributes disproportionately to the power that leaks into the EoR window.

The array configuration has a number of advantages that we would like to preserve:
\begin{enumerate}
\item Its compactness enables it to achieve a high -sensitivity detection of the 21\,cm signal using only foreground avoidance. 
\item It can be calibrated precisely using its redundancy.
\item Its hexagonal configuration makes FFT correlation possible \citep{FFTT, FFTT2}, although it is not strictly necessary for an array with only a few hundred elements.
\end{enumerate}
It also has a number of drawbacks:
\begin{enumerate}
\item It has fairly low angular resolution. 
\item It measures only a relatively small number of modes simultaneously---eventually, sample variance dominates these measurements. \item The modes it does measure fall on a regular hexagonal grid, leading to grating lobes in the point spread function (PSF). 
\end{enumerate}

In this section, we identify strategies for mitigating these shortcomings with minor modifications to the array configuration. In particular, we asses methods of subsampling the hexagonal baseline grid (Section \ref{sec:subsampling}) and for increasing the angular resolution of the array by adding far-flung ``outrigger'' antennas (Section \ref{sec:increaseRes}). Then, we show that all these designs have approximately the same raw power spectrum sensitivity.

\subsection{Subsampling the Baseline Grid} \label{sec:subsampling}

Although a 331-element hexagonal core measures 54,615 visibilities at any one time, it actually only measures 630 unique separations between antennas. These fall on a hexagonal grid, since the convolution of a hexagon with itself is simply a larger hexagon. In a compact array, the minimum separation between antennas, 14.6\,m, sets the minimum baseline length and the fundamental size of that grid of measured baselines. And since the Fourier sky is regularly sampled at spacings much wider than Nyquist sampling ($\lambda/2$), we observe grating lobes in the PSF due to signal aliasing caused by undersampling. If we want to suppress grating lobes, we need denser sampling and more independent information.

Of course, our visibility measurements are not disconnected \mbox{$\delta$-functions} in Fourier space. The visibility, which is a measurement of the correlation of voltages measured by two antennas separated by baseline vector $\base$, is given by 
\begin{align}
V(\base,\nu) = \int B\left(\shat,\nu \right)
I(\shat,\nu) \exp\left[-2 \pi i  \frac{\nu}{c} \base \cdot \shat \right] d\Omega . \label{eq:AnalyticVisibility}
\end{align}
where $I(\shat,\nu)$ is the sky intensity as a function of position and frequency and $B(\shat,\nu)$ is the product of the direction and frequency dependence of the response of individual antennas.\footnote{For which, in this work, we use a simulation of actual HERA elements (\citealt{HERADish1,HERADish2,HERAOverview}).} 

For a coplanar array sampling the sky instantaneously, we can think of a visibility as a average over a region of the Fourier transformed sky, projected onto a plane and convolved with the Fourier transform of the beam product \citep{ThompsonMoranSwenson}. Alternatively, we can think of it as a \mbox{$\delta$-function} sampling of the beam-convolved sky. HERA's beams are roughly axisymmetric and therefore our weighted average is over an approximately circular region of Fourier space (also known as the $uv$-plane). It follows that we have the least information about the parts the $uv$-plane furthest from the exact baselines probed. If we would like to make new measurements inside the core (in $uv$-space) as independent as possible (although not, of course, completely independent of existing baselines), then we would like to place antennas so as to produce new baselines at the centroids of the lattice of equilateral triangles that make up the hexagonal grid.

\begin{figure}[] 
	\centering 
	\includegraphics[width=.48\textwidth]{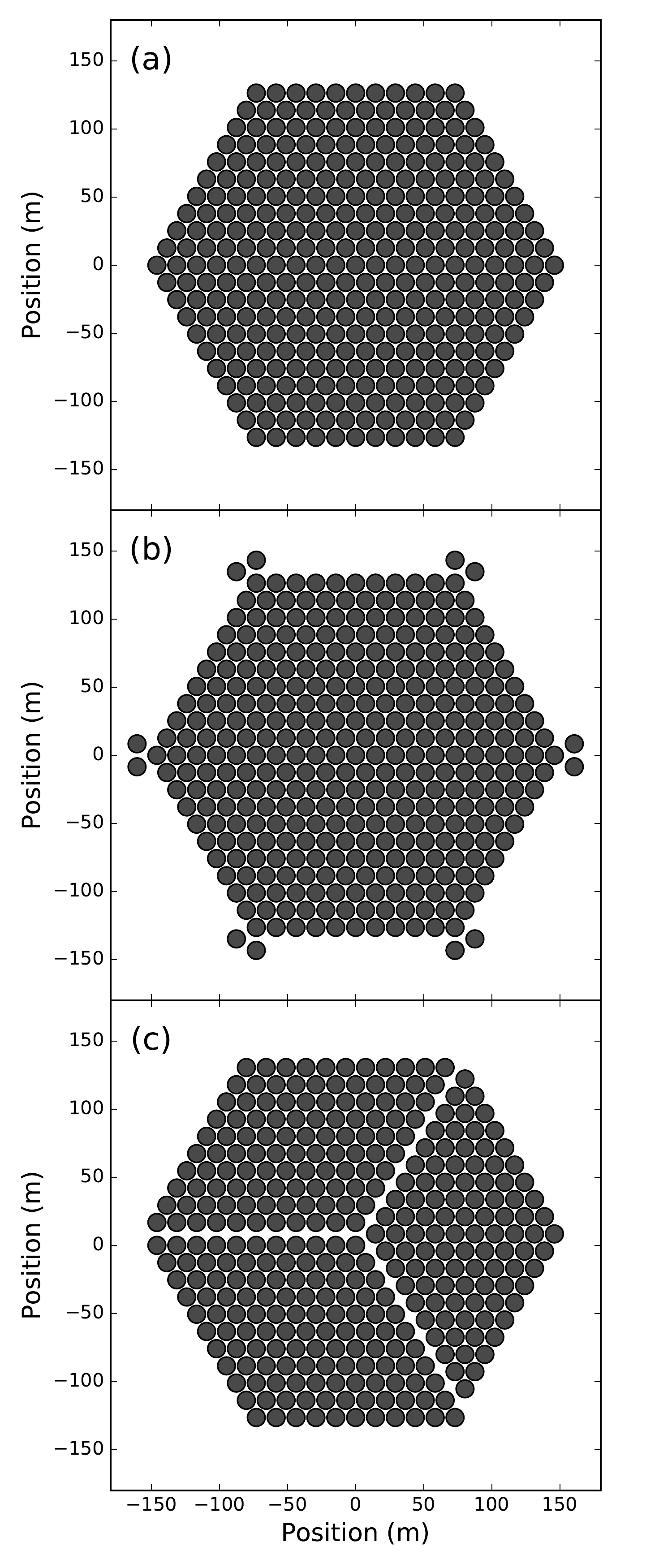}
	\caption{Three different versions of the HERA core, meant to produce different instantaneous baseline distributions. In (a) we take the standard hexagon, which produces a hexagonal grid of baseline separations (see Figure \ref{fig:redundancy}). In (b) we add pairs of antennas that are displaced $1/3$ or $2/3$ of the sum of two hexagonal basis vectors (e.g.\ right and up-and-to-the-right). In (c) we fracture the entire array into three sectors, displaced $0/3$, $1/3$, or $2/3$ of the sum of two hexagonal basis vectors. The configurations have 331, 343, and 331 antennas, respectively. These different options affect both mapmaking and calibratability, as we discuss in Sections \ref{sec:mapmaking} and \ref{sec:calibration}.}
	\label{fig:arrays}
\end{figure}

\begin{figure*}[] 
	\centering 
	\includegraphics[width=.98\textwidth]{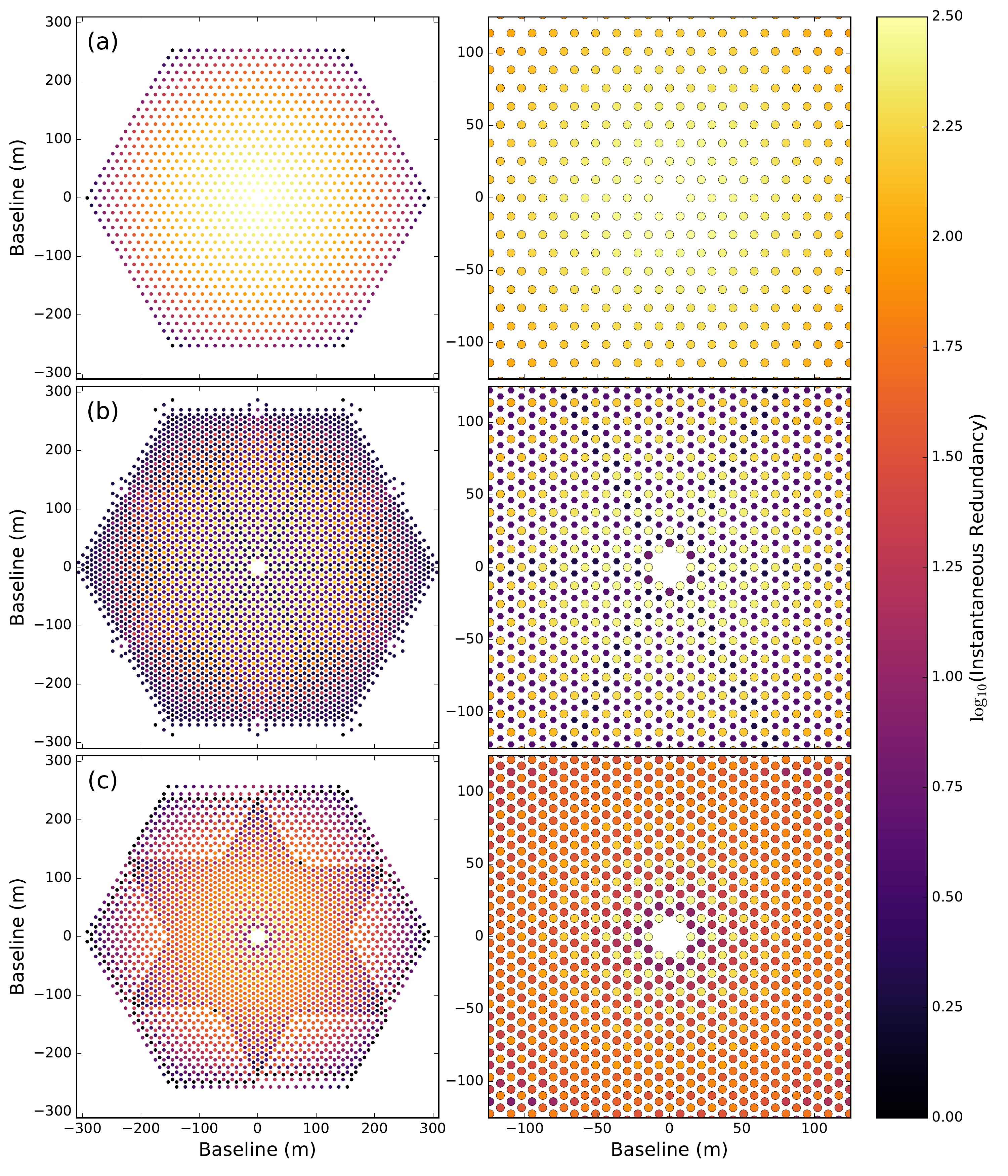}
	\caption{The baseline distributions produced instantaneously by the three different array configurations in Figure \ref{fig:arrays}. Color indicates how many antenna pairs measure the given separation vector at any one time. On the left we show the full baseline pattern, on the right we zoom into the highest sensitivity region. In the right panels, baselines sampled four or fewer times are indicated by smaller markers in order to better illustrate the redundancy pattern. An unmodified core in (a) produces a hexagonal grid of well-sampled baselines. Configuration (b) adds off-grid antennas to each corner of the core to produce baselines that effectively subsample the hexagonal grid. The splitting of the core in (c) achieves the same sampling pattern as in (b) without any additional antennas. In contrast to (b), the off-grid baselines in (c) are more frequently and evenly sampled. That is because longer baselines are usually between relatively displaced sectors. This also leads to unsampled points on the grid outside the main hexagram, but that deficiency can be rectified with outrigger antennas (see Figure~\ref{fig:fullRedundancy}). The size of the points is not physical; individual baselines measure \mbox{$\delta$-function} samples of the beam-convolved sky. Therefore, the sub-aperture samples of configurations (b) and (c) are not measuring completely independent sky information from their neighboring baselines.}
	\label{fig:redundancy}
\end{figure*}

For the sake of comparison, we start with a ``fiducial'' HERA core, which we call configuration (a) and show in the first panel of Figure \ref{fig:arrays}.
In Figure \ref{fig:redundancy} we can see that hexagonal grid pattern of instantaneously redundant baselines (as opposed to baselines that measure the same mode at a later time due to the rotation of the earth.) 
We note that the best-sampled baselines are the shortest. Longer separations have fewer corresponding pairs of antennas within the core.

To produce new baselines at the centroids of the triangular lattice, we need to add antennas that are not on the hexagonal grid or dither existing antennas. If we pick any two 14.6\,m basis vectors that define the hexagonal grid in linear, integer combinations (e.g.\ up-and-to-the-left and up-and-to-the-right), we note the centroids fall at 1/3 or 2/3 of the sum of those two vectors. In other words, by introducing antennas dithered from the main grid by 1/3 or 2/3 of that separation, we can add a grid of baselines that effectively subsample the main hexagonal grid. Maximally packed dishes Nyquist sample modes within the main lobe of the primary beam, but they produce grating lobes in the sidelobes (which would require $\lambda/2$ spacing to eliminate for the whole sky). Sub-aperture sampling suppresses these grating lobes, as we will see in Section \ref{sec:mapmaking}.

We employ this strategy in developing our two alternative array configurations, (b) and (c), shown in Figure \ref{fig:arrays}. In (b), we add a small number of antennas near the perimeter of the main hexagon, each displaced from the main grid. In (c) we split the entire hexagon into three sectors, each relatively displaced by those same increments. Both patterns allow for complete sampling at triple the density of the original grid (see Figure \ref{fig:redundancy}). In the case of the split-core configuration (c), that triple-density coverage does not extend to the full hexagonal core, but those baselines can be filled in by outrigger antennas, as we will discuss in Section \ref{sec:increaseRes}.\footnote{As an aside, the bottom row of (c) is not necessary for the desired $uv$-coverage and can be eliminated as a cost saving measure. In this paper, we keep it to make the cores more directly comparable.}

An important contrast between configurations (b) and (c) is that while off-grid baselines in (b) have at most a handful of measurements, the distinction between ``on-grid'' and ``off-grid'' in (c) becomes less clear for longer baselines. That is because longer baselines are more likely to be inter-sector and thus off-grid while shorter baselines are more often intra-sector and thus on-grid. The result is more distributed and thus uniform coverage of the triple-density grid than in (b).\footnote{We considered carrying this further and splitting the hexagon into nine sectors to create even finer gridding. It turns out that it is difficult to make such a configuration with full $uv$-coverage and they appear to be less redundantly calibratable (and probably more difficult to build). We therefore do not consider such a design in this work.} It should be noted that configurations (a) and (b), by virtue of their denser cores, concentrate somewhat more sensitivity on shorter baselines, which are those less affected by the wedge. We also note that (b) and (c) do not substantially affect the possibility of performing FFT correlation, since it merely requires that all antennas fall on a regular or hierarchically regular grid \citep{FFTT2}; gridding based FFT schemes \citep{moff,moff2} are not necessary. That grid is now simply finer and has many unobserved vertices that must be flagged.

\subsection{Increasing Angular Resolution} \label{sec:increaseRes}

\begin{figure}[] 
	\centering 
	\includegraphics[width=.48\textwidth]{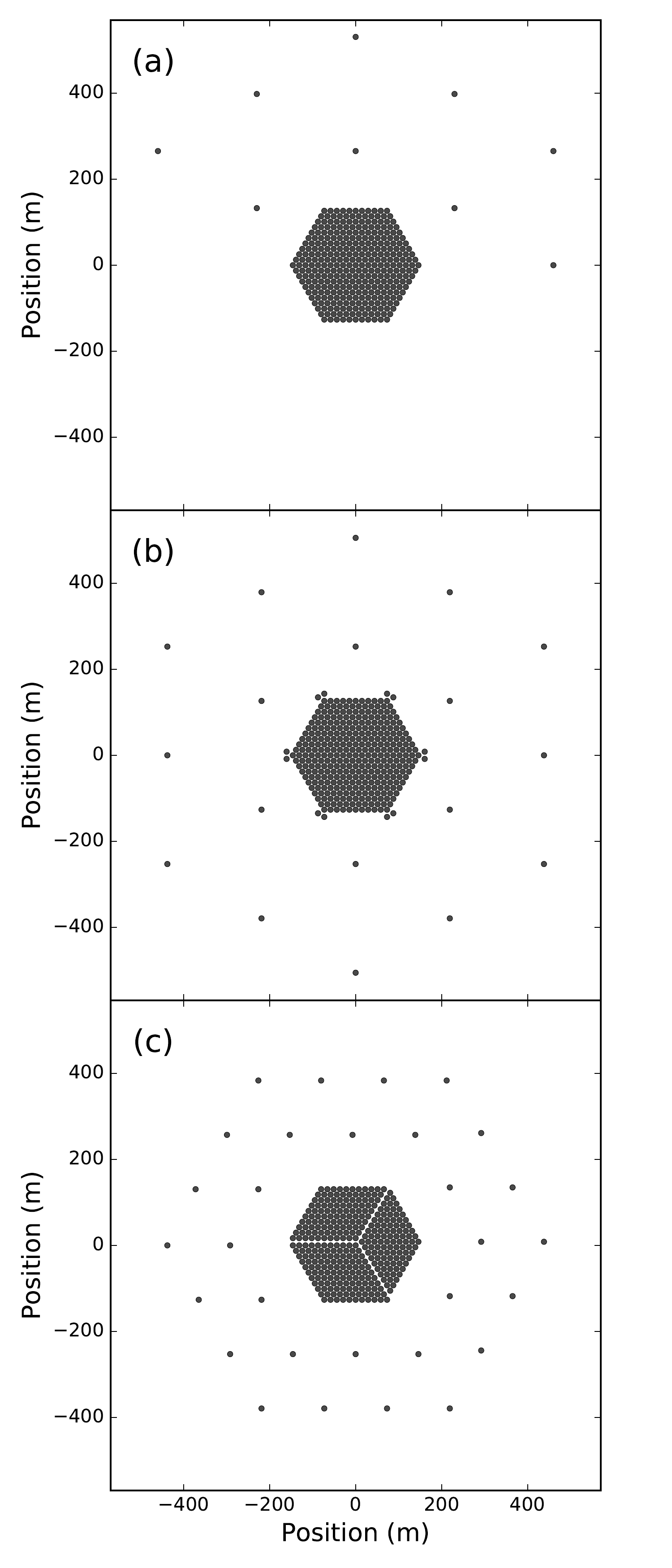}
	\caption{Three different approaches to ``outrigger'' antennas that increase the angular resolution of the array core designs in Figure \ref{fig:arrays}. The outrigger antennas in (a) and (b) are designed to tessellate hexagonal sampling regions formed by outrigger-core antenna pairs in order to efficiently maximize the region of full coverage and thus increase angular resolution. However, (b) includes antennas on both sides of the core and is slightly more compact than (a), enhancing redundant calibratability (see Section \ref{sec:calibration}). By contrast, the outrigger antennas in (c) are designed for both redundant calibration and to extend triple-density $uv$-coverage to longer baselines (see Figure \ref{fig:fullRedundancy}). They have 340, 361, and 361 antennas, respectively.}
	\label{fig:fullArrays}
\end{figure}

\begin{figure*}[] 
	\centering 
	\includegraphics[width=.98\textwidth]{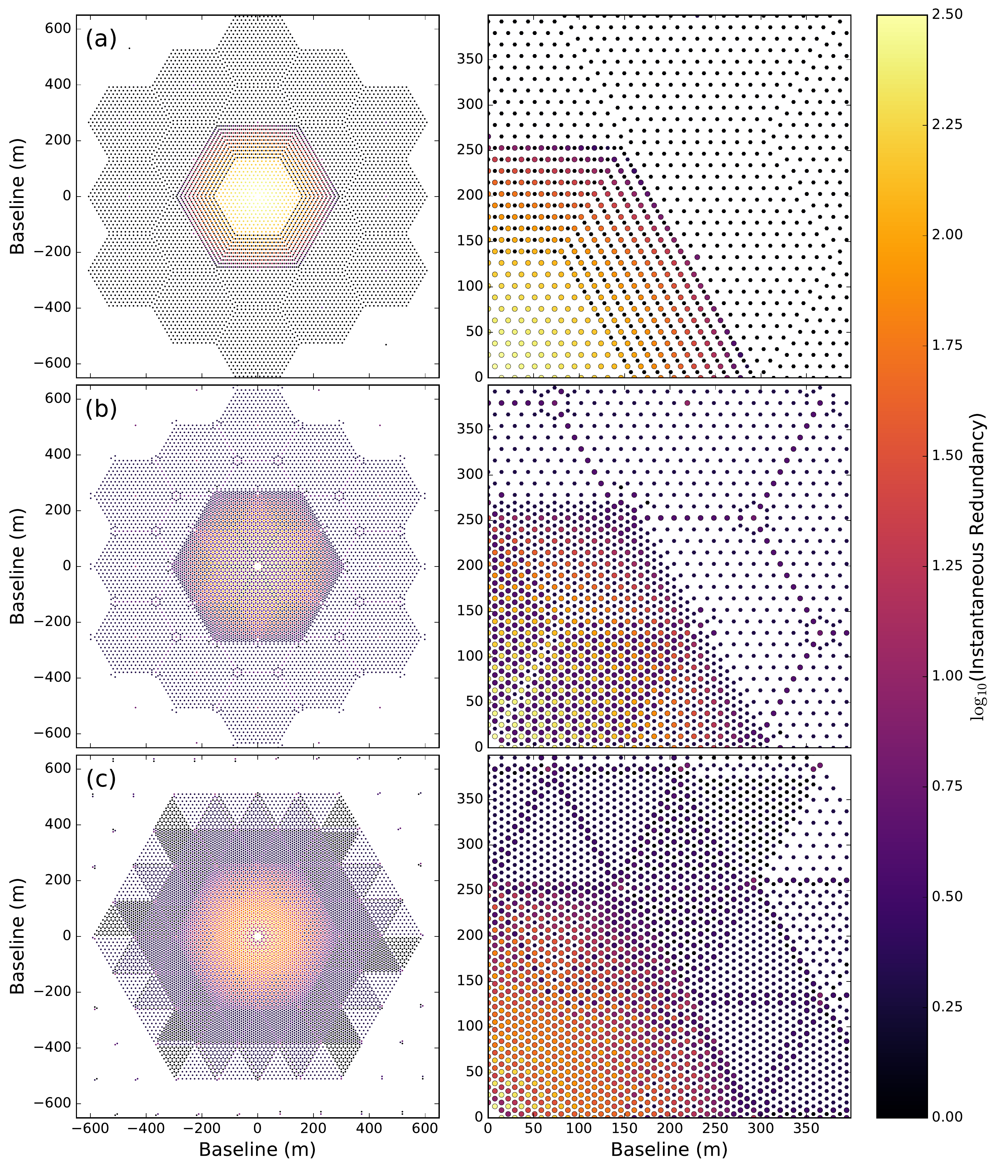}
	\caption{Instantaneous baseline sampling produced by the three different outrigger configurations in Figure \ref{fig:fullArrays}. On the left we show the full baseline sampling pattern. On the right we show a zoom into a representative section. Here baselines sampled two or fewer times are denoted by smaller markers in the righthand panels in order to better illustrate the network of redundancy. In particular, we can see how contracting the outrigger positions from configuration (a) to configuration (b) produces overlapping edges of redundantly sampled baselines, creating a network by which the whole array can be redundantly calibrated (see Section \ref{sec:calibration}). By contrast, the design in (c) arranges baselines to achieve complete coverage at triple density to as large a radius as possible while also maintaining plenty of overlap for redundant calibration. Additional outrigger antennas can be added to produce expanded $uv$-coverage for higher angular resolution imaging.}
	\label{fig:fullRedundancy}
\end{figure*}

In addition to the hexagonal cores, we are interested in supplementing each design with several far-flung ``outrigger'' antennas in order to improve angular resolution (see Figure \ref{fig:fullArrays}). Adding a single antenna far from the core adds an entire hexagon of baselines to the $uv$-plane.
This improves the imaging capability of the array and can help with foreground removal (see Section \ref{sec:mapmaking}). With configuration (a), we pick the minimal set of additional antennas to tile the maximum area of the $uv$-plane with hexagons. Although hexagons do tessellate, the borders between them create discontinuities in the hexagonal grid, as we can see in Figure \ref{fig:fullRedundancy}.
Tessellating hexagons produces some overlapping regions in the core, but outside the core the scheme is extensible to longer baselines with no additional overlap.

Configuration (b) takes inspiration from configuration (a) by also using the hexagonal perimeter of the core to tile the $uv$-plane. Its outrigger antennas differ in two notable ways. First, it includes outrigger antennas on both sides of the core. Second, that grid is contracted slightly from the grid in (a) such that there exist redundant baselines measured between different outrigger antennas. Both these changes are designed to allow for redundant calibratability, which we will discuss in Section \ref{sec:calibration}.

Finally, configuration (c) takes a different approach. While it too is designed for redundant calibratability, it takes advantage of the split-core configuration to create full baseline coverage at triple density (see Figure \ref{fig:fullRedundancy}). Compared to (a) or (b), that increased density allows it to better suppress the sidelobes of point sources at the cost of some angular resolution.

\subsection{Raw Power Spectrum Sensitivity}

Before we move on to compare these configurations in terms of mapmaking and calibration, it is useful to check that these designs do not differ significantly in terms of their raw sensitivity to the 21\,cm power spectrum. Since most sensitivity comes from short baselines and all three configurations have very similar cores, we do not expect much variation.

A proper calculation of the significance with which a fiducial model of reionization can be detected by these arrays is quite difficult. In addition to depending on foreground removal strategies \citep{PoberNextGen} and models for the covariance of foreground residuals \citep{EmpiricalCovariance}, one should take into account the partial coherence of baselines that measure overlapping modes in the $uv$ plane, either from the same baseline at different times or from sub-aperture sampling. This is possible, though computationally challenging. No algorithm exists yet that propagates these effects into the power spectrum \citep{mapmaking}. A full treatment is therefore outside the scope of this paper.

Instead, we follow \citet{pober_hera4} and employ the \texttt{21cmSense} code\footnote{First developed for \citet{PoberNextGen} and now available publicly at \url{https://www.github.com/jpober/21cmSense}.} to explore the sensitivities of three configurations. This is quantified in terms of a ``cumulative detection significance,'' which is the square root of the sum in quadrature of the ratio of the 21\,cm power spectrum to the thermal noise over all wavenumbers $k$. In particular, we look at an 8\,MHz bandwidth centered on 135\,MHz for a 1080\,hour drift scan and calculate the sensitivity to a model power spectrum from \citet{21cmFast} with a midpoint of reionization at $z=9.5$. Table \ref{tab:SNR} shows that, regardless of our foreground mitigation strategy, all three of our arrays perform very similarly. We therefore should consider other criteria to assess their relative merit.
\begin{table}[t]
  \centering
\begin{ruledtabular}
\caption{Estimated detection significance for the various configurations.\label{tab:SNR}}
  \begin{tabular}{ c c c c  c  }
\textbf{Configuration} & $N_\text{ant}$ & $N_\text{core}$  &  \textbf{Mod.~S/N} & \textbf{Opt.~S/N}\\ \hline
Fiducial (a) & 340 & 331 & 24.4$\sigma$ & 91.9$\sigma$ \\ 
Corners (b) & 361 & 343 & 25.0$\sigma$ & 95.2$\sigma$ \\ 
Split-core (c) & 361 & 331 & 24.0$\sigma$ & 93.7$\sigma$ \\ 
  \end{tabular}
\end{ruledtabular}
\tablecomments{This calculation is performed for a reionization model that reaches 50\% ionization at $z=9.5$ \citep{21cmFast} for each array configuration. We examine both a moderate scenario where foregrounds must be avoided outside the horizon window (plus a $0.1$\hiMpc buffer), and an optimistic scenario where foregrounds outside the primary beam wedge can be subtracted. This approximate treatment using \texttt{21cmSense} does not account for partial coherence between baselines that sample overlapping regions of the $uv$ plane, and therefore affords no advantage for the off-grid baseline sampling of configurations (b) and (c). Instead the results are dominated by the number antennas, $N_\text{ant}$, and (especially for the moderate foreground scenario) by their relative compactness (which is related to $N_\text{core}$). These relatively small differences in sensitivity are unlikely to impact suitability for 21\,cm cosmology.}
\end{table}


\section{Mapmaking Capability} \label{sec:mapmaking}
The modifications and additions to the core HERA configuration that we presented in Section \ref{sec:design} were largely motivated by the desire to increase the density and extent of baseline sampling of the $uv$-plane in order to supplement HERA's imaging capability. It is worthwhile therefore to assess the ways in which imaging capabilities are different between the different configurations. As we alluded to earlier, we expect the outrigger antennas and off-grid antennas to affect angular resolution and widefield imaging, respectively. We can assess these effects qualitatively by comparing PSFs, which we do in Section \ref{sec:PSFs}. We can also be more quantitative in our comparison by looking at the expected noise and noise correlations in the final maps, as we do in Section \ref{sec:mapmakingResults}. But first, we begin in Section \ref{sec:mapReview} with a review of the optimal mapmaking formalism of \citet{mapmaking} that underlies the rest of the detailed mapmaking comparisons in this work.

\subsection{Mapmaking Review} \label{sec:mapReview}

Fundamentally the problem of mapmaking is one of data reduction. We have a large quantity of time-ordered measurements, expressed as a data vector $\y$, which sample different linear combinations of the sky at different times and in different ways. However, those data are noisy, so we must be cognizant of how we weight and combine them.

The action of the interferometer measuring visibilities (as in Equation \ref{eq:AnalyticVisibility}) can be expressed as a relationship between some discretized sky $\x$ and our measurements $\y$ as
\beq
\y = \A \x + \n \label{eq:measurement}.
\eeq
Here $\A$ represents the interferometric measurement---the family of integrals like Equation \ref{eq:AnalyticVisibility} but discretized---and $\n$ is the thermal noise. The discretization of $\A$ is demonstrated in \citet{mapmaking}, which also includes a more detailed discussion of the following definitions and derivations. 

If the statistics of the thermal noise are Gaussian and we define $\N \equiv \langle \n \n^\dagger \rangle$, then the optimal estimator (in the sense that no information about the sky is lost), $\xhat$, is 
\beq
\widehat{\x} = \D \A^\dagger \N^{-1} \y \label{eq:OMM}
\eeq
where $\D$ is some invertible normalization matrix \citep{TegmarkCMBmapsWOLosingInfo}. To understand the statistics of this estimator, we should look at its mean and covariance. Since the sky is not random and since the thermal noise has $\langle \n \rangle = 0$, 
\begin{align}
\langle \widehat{\x} \rangle &= \langle \D \A^\dagger \N^{-1} (\A \x + \n) \rangle \nonumber \\ 
&= \D \A^\dagger \N^{-1} (\A \x + \langle \n \rangle) \nonumber \\ 
&= \D \A^\dagger \N^{-1} \A \x = \PSF \x, \label{eq:<xhat>}
\end{align}
where we have defined 
\beq
\PSF \equiv \D \A^\dagger \N^{-1} \A. \label{eq:PSFdef}
\eeq
The matrix $\PSF$ is the matrix of PSFs that tells us how a source in any pixel in $\x$ will appear in all other pixels. Likewise the covariance of the estimator is given by 
\begin{align}
\C &= \left< (\widehat{\x} - \langle\xhat\rangle) (\widehat{\x} - \langle\xhat\rangle)^\dagger \right> \nonumber \\
&= \left<(\D\A^\dagger\N^{-1}\n)(\D\A^\dagger\N^{-1}\n)^\dagger\right> \nonumber \\
&= \D\A^\dagger\N^{-1}\left< \n \n^\dagger\right> \N^{-1} \A \D^\dagger \nonumber \\
&= \PSF \D^\dagger \label{eq:CN}.
\end{align}
Since the only random thing here is the noise, this $\C$ is really the noise covariance of every map pixel with every other map pixel.

For estimating power spectra, \citet{mapmaking} argued that choosing a simple diagonal form of $\D$ is ideal for propagating mapmaking statistics into the power spectrum. In this work, we simply take $\D \propto \Eye$ such that the PSF of a source at zenith peaks at 1. In the flat-sky approximation used by most radio interferometers with small fields of view, this choice is analogous to natural weighting, where every independent measurement is weighted only by the noise in that measurement and not by the number of other measurements that fall in the same $uv$-cell.

However, it is also useful to consider the other extreme choice for $\D$. If we wanted an unbiased estimator of the sky, one where $\langle \xhat \rangle = \x$, then we should choose $\D = (\ANA)^{-1}$ in order to obtain $\PSF = \Eye$. In the flat-sky approximation, this choice is analogous to uniform weighting of cells in the $uv$-plane. With a limited field of view (low $uv$ resolution) and limited angular resolution (small maximum $u$ and $v$), it is possible to sample every $uv$-cell and produce a \mbox{$\delta$-function} PSF using uniform weighting. This generally produces very noisy maps. And of course, depending on the pixelization, $\ANA$ is often not invertible and this procedure becomes impossible without further assumptions about the sky. However, when this procedure is possible, it follows then that 
\beq
\C = \PSF \D^\dagger \longrightarrow (\ANA)^{-1}.
\eeq
Since inverse covariance weighting for a future data reduction step involves multiplication by $\C^{-1} = \ANA$, this means that $\ANA$ measures the information content in our maps. In Section \ref{sec:mapmakingResults}, we will explore the structure of this matrix for our array configurations.

\subsection{PSF Comparison} \label{sec:PSFs}

\begin{figure*}[] 
	\centering 
	\includegraphics[width=1\textwidth]{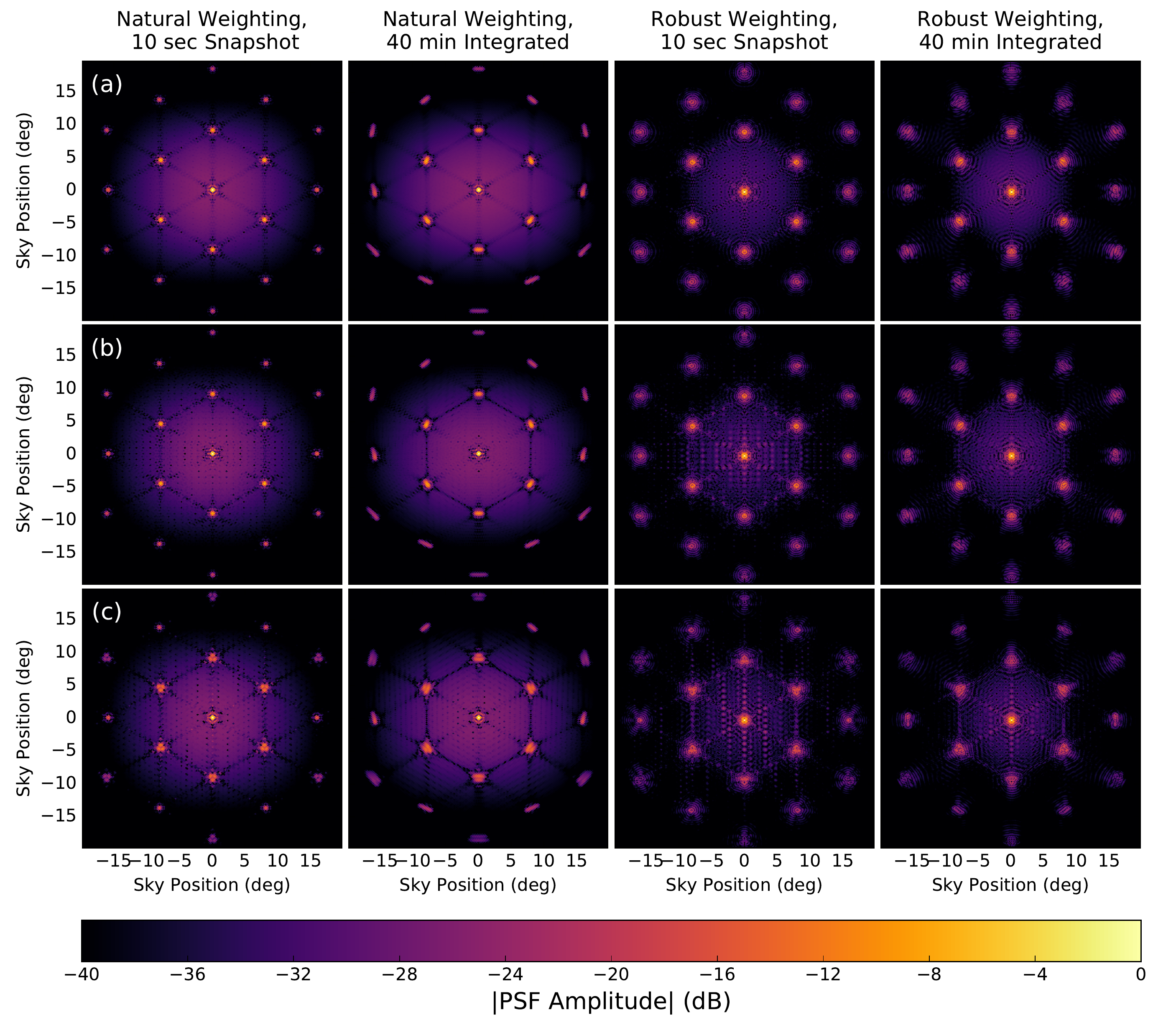}
	\caption{Point spread functions for a source at zenith, imaged at 150\,MHz with all three array configurations in Figure \ref{fig:fullArrays}. We show both single 10 second snapshots and full 40 minute integrations, corresponding to the passage of the field center through the primary beam. As we would expect, the sidelobes of the PSF or ``synthesized beam'' are somewhat suppressed and smeared out in configuration (c). The effect is small in the naturally weighted PSFs, since most of the sensitivity in all configurations is concentrated on the baselines in the hexagonal grid created by the core. We also include a ``robust'' weighting scheme, where no baseline is weighted more than 50 times more heavily than any other, in order to further highlight the sidelobe suppression that configuration (c) achieves with its relatively equitable distribution of sensitivity between on-grid and off-grid baselines.}
	\label{fig:highResPSFs}
\end{figure*}

We start with a more qualitative comparison of the PSFs of the three configurations. In Figure \ref{fig:highResPSFs}, we examine the central region of a high-resolution HEALPix \citep{healpix} map of the PSF  of a source in the center of a $20^\circ$ facet ($N_\text{side} = 256$). At 150\,MHz, that $20^\circ$ corresponds to roughly double the size of HERA's primary beam.
We examine the PSF produced by a single 10 second snapshot with the facet center at zenith and by a 40 minute drift scan observation centered on the facet. We also examine two weighting schemes, one where $\N$ simply reflects noise on a given redundantly sampled baseline (e.g.\ ``natural'' weighting) and another where no redundant baseline is weighted more than 50 times higher than any other. This is akin to ``robust'' weighting \citep{ThompsonMoranSwenson}, which transitions from natural weighting of low signal-to-noise ratio (S/N) baselines to uniform weighting of high S/N baselines, balancing noise against sidelobe confusion. 

At first glance, the PSFs are very similar. This is to be expected, especially for the naturally weighted PSFs, which are so dominated by the core baselines. The prominent six-fold sidelobes we see in all panels of Figure \ref{fig:highResPSFs} are due to the inter-element spacing and configuration in the core.

There are some small differences between the configurations. The main lobe of the PSF is slightly more sharply defined in configurations (a) and (b) because their $uv$-coverage extends to longer baselines (see Figure \ref{fig:fullRedundancy}). More prominently, the sidelobes are dimmer and more smeared out in configuration (c) than in either (a) or (b). This makes sense since we expected the sidelobes to be suppressed in the split-core configuration, which gives more weight to the off-grid baselines. Our ``robust'' weighting scheme helps us see this more clearly, since it gives relatively more weight to off-grid baselines and better demonstrates the sidelobe suppression possible with configuration (c).

\subsection{Quantitative Mapmaking Results} \label{sec:mapmakingResults}

The observed sidelobe suppression in Figure \ref{fig:highResPSFs} raises a question: does configuration (c) actually have an improved ability to measure modes far from the center of the primary beam ``more independently''? To answer that, we need to analyze the structure of the matrix $\ANA$, the noise-weighted instrument response which contains the information and correlations in our map. Were it computationally feasible, we would do this for a full-sky, high-
resolution map. However, the numerical difficulty of applying the rigorous mapmaking treatment of Section \ref{sec:mapReview} necessitates some approximations \citep{mapmaking}. 

\subsubsection{Widefield Mapmaking}

\begin{figure*}[] 
	\centering 
	\includegraphics[width=1\textwidth]{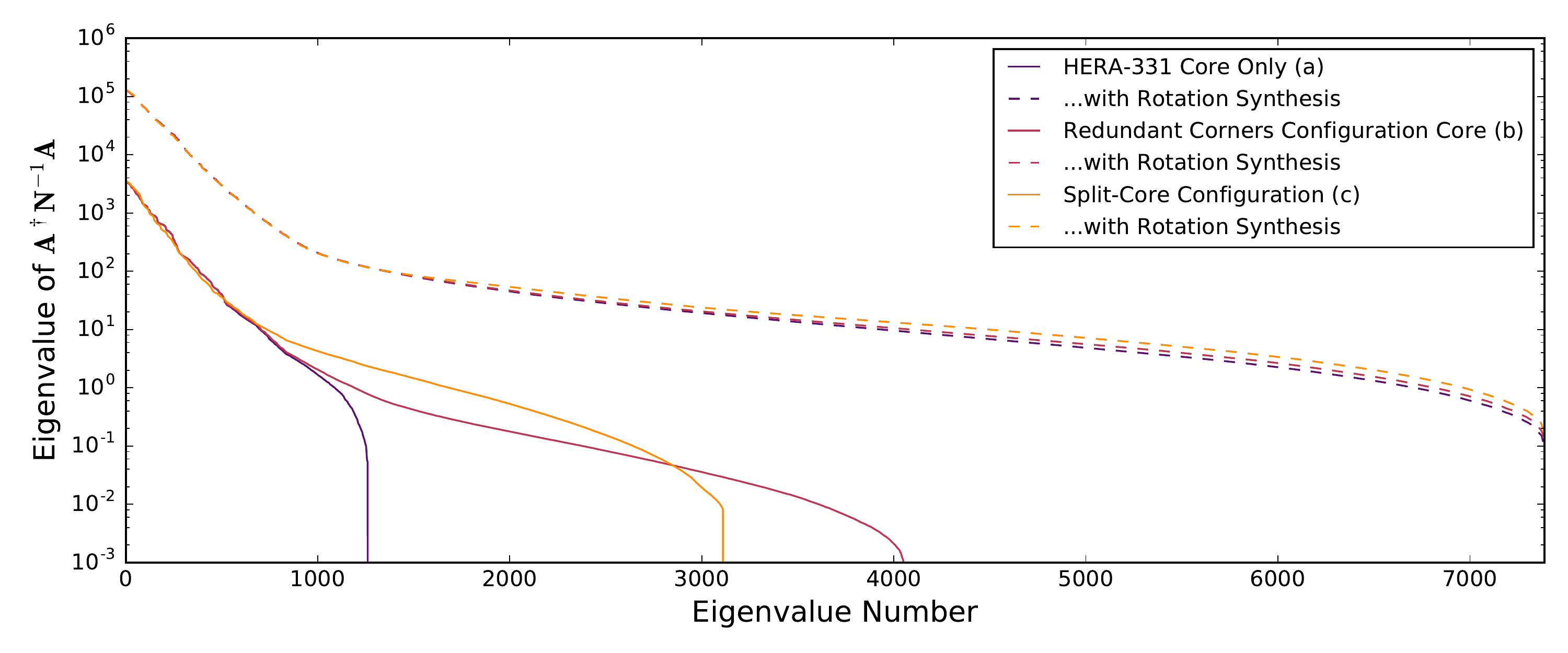}
	\vspace{-5pt}
	\caption{To assess the instantaneous and time-integrated widefield imaging capabilities of the array cores in Figure \ref{fig:arrays}, we calculate the eigenvalues of $\ANA$, the noise-weighted instrument response, for each. The amplitude of the eigenvalues of this matrix tells us how well different modes in the map are measured. It is also the matrix we would have to invert if we want to set the PSF to the identity. In this case, we examine a dynamically pixelized sky designed with HEALPix $N_\text{side} = 64$ inside the main lobe of the primary beam and decreasing resolution at lower altitudes to maintain roughly constant beam power per pixel. Results are rescaled by the number of antennas to make the comparison fair; the units are arbitrary. We find that the addition of dithered antennas and off-grid baselines improves instantaneous mapmaking considerably, but the difference between configurations after Earth rotation synthesis (which of course adds information) is slight. Configuration (a), the unmodified hexagonal core, measures the fewest modes instantaneously. Configuration (b) can measure more independent modes than configuration (c), but with less sensitivity. This makes sense given the calculated instantaneous redundancy (Figure \ref{fig:redundancy}).}
	\label{fig:fullSkyEVs}
\end{figure*}

First we focus on how the core configuration (see Figure \ref{fig:arrays}) affects widefield mapmaking. We therefore plot the eigenvalue spectra of all three array configurations in Figure \ref{fig:fullSkyEVs}.
In order to overcome the numerical difficult of evaluating $\ANA$ and its eigenspectrum, we pixelize the sky adaptively, using the hierarchical property of HEALPix to develop a pixelization scheme that roughly preserves equal primary beam power per pixel. This scheme ranges from $N_\text{side} = 64$ around zenith to $N_\text{side} = 4$ near the horizon. We perform this analysis for both a single snapshot and 40 minutes of rotation synthesis. We then calculate the eigensystem, normalizing by the number of antennas to make the comparison fair.

At first glance, the eigenvalues---which encode the amount of information in each mode---are very similar, especially after Earth rotation synthesis. The biggest differences are revealed only in the case of instantaneous imaging, where we see a sharp cutoff for configuration (a) and a factor of a few difference between (b) and (c) in some of the more poorly measured eigenmodes.\footnote{To be clear, each array corresponds to a different $\ANA$ matrix. Although their eigenmodes are similar, it is misleading to directly compare the $N$th eigenvalue without also considering the modes they correspond to.} We expect that the performance of configuration (c) is due to the more even distribution of on-grid and off-grid baselines (see Figure \ref{fig:redundancy}).

\begin{figure*}[p] 
	\centering 
	\includegraphics[width=.85\textwidth]{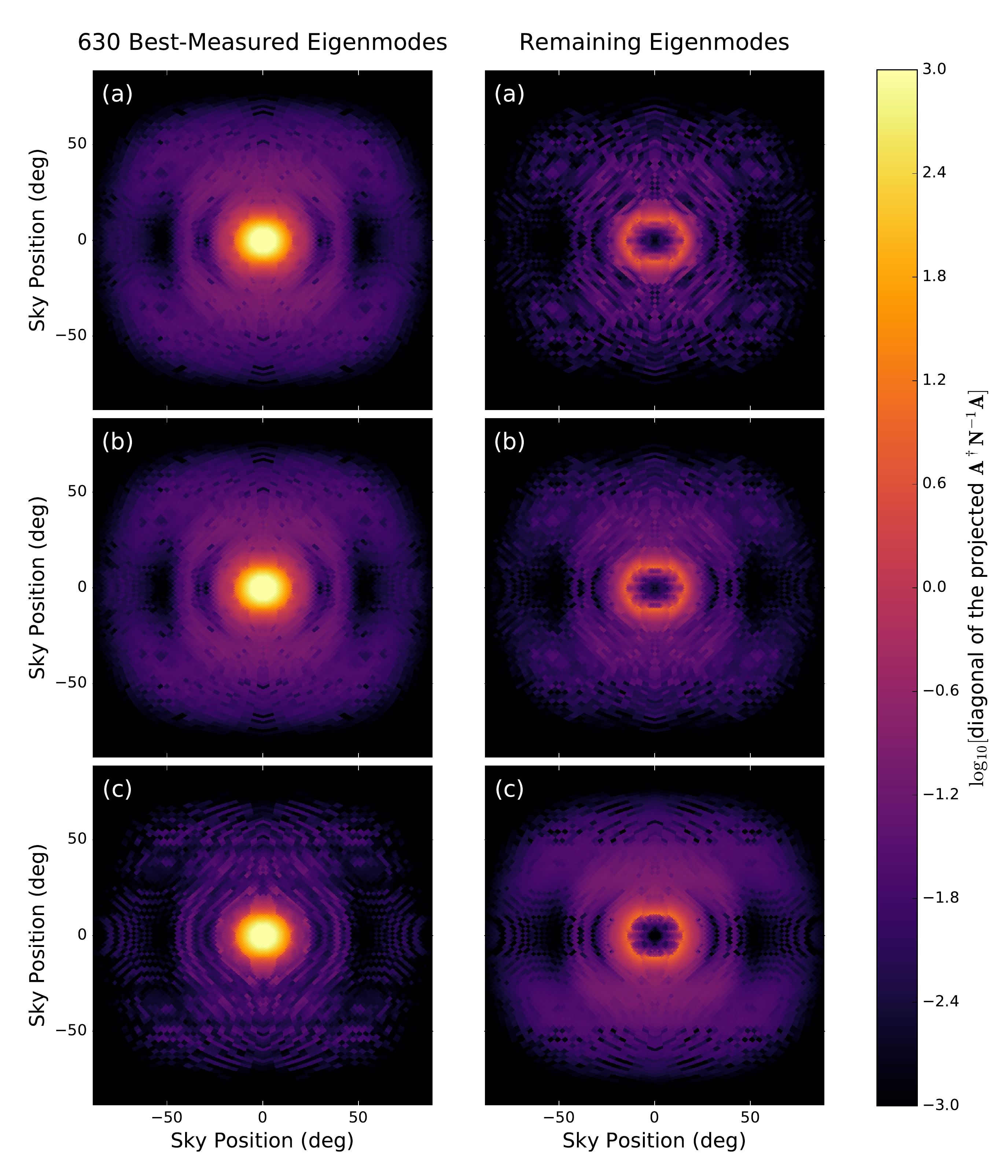}
	\caption{We can better understand the relative snapshot mapmaking performance of the array configurations in Figure \ref{fig:fullSkyEVs} by looking at the diagonal elements of $\ANA$ (which is also the inverse noise covariance of the array) after it's been split into two orthogonal subspaces---one the subspace spanned by the eigenvectors corresponding the 630 largest eigenvalues, the other by the rest of the eigenvectors. 630 is where the split-core configuration (c) begins to outperform the others (see Figure \ref{fig:fullSkyEVs}). While the effect is modest, we see that the measurements of configuration (c) exhibit a greater separation between modes in the main lobe of the primary beam and modes in the sidelobes. This can be understood as a consequence of the brighter grating lobes of the PSF we saw in Figure \ref{fig:highResPSFs} for configurations (a) and (b), which correlate modes inside and outside the main lobe of the primary beam.}
	\label{fig:fullSkyProjected}
\end{figure*}

To verify our intuition about the meaning of these eigenvalue spectra, we should look at the eigenvectors of the single snapshot. It is difficult to interpret or compare individual eigenvectors. Instead, we want to look at each $\ANA$ matrix in two different subspaces---one corresponding to the first 630 eigenvalues where, in Figure \ref{fig:fullSkyEVs}, the spectra are very similar and the other corresponding to the rest where they differ the most. In Figure \ref{fig:fullSkyProjected} we show the diagonals of $\ANA$ for each array configuration, projected into the two orthogonal subspaces.
One can think of these matrices as $\ANA$ where either the first 630 or the rest of the eigenvalues are set to zero. 

Roughly speaking, Figure \ref{fig:fullSkyProjected} confirms our intuition that the highest information modes are those concentrated in the main lobe of the primary beam. That makes sense, since that is where the telescope is most sensitive. However, when we compare configuration (c) to either (a) or (b), we see a much clearer separation of information. The modes that contain information about the widefield are not the same modes as those that contain information about the primary field of view. It is not that configuration (c) is more sensitive to the modes deep in the primary beam. In fact, when when we add these two projected matrices back together, the results for all three arrays are nearly identical. Rather, what this result tells us is that we can make maps that more easily separate sources in the sidelobes from signal in the main lobe of the primary beam. As \citet{NithyaPitchfork} and \citet{PoberSidelobe} argue, these are the sources most likely to leak into the EoR window. The exact level of that isolation depends both on particular field of view and the observing frequency, both of which set the relative contributions of foregrounds in the main lobe and in the sidelobes of the primary beam.

\subsubsection{High-resolution Mapmaking}

\begin{figure*}[t!] 
	\centering 
	\includegraphics[width=1\textwidth]{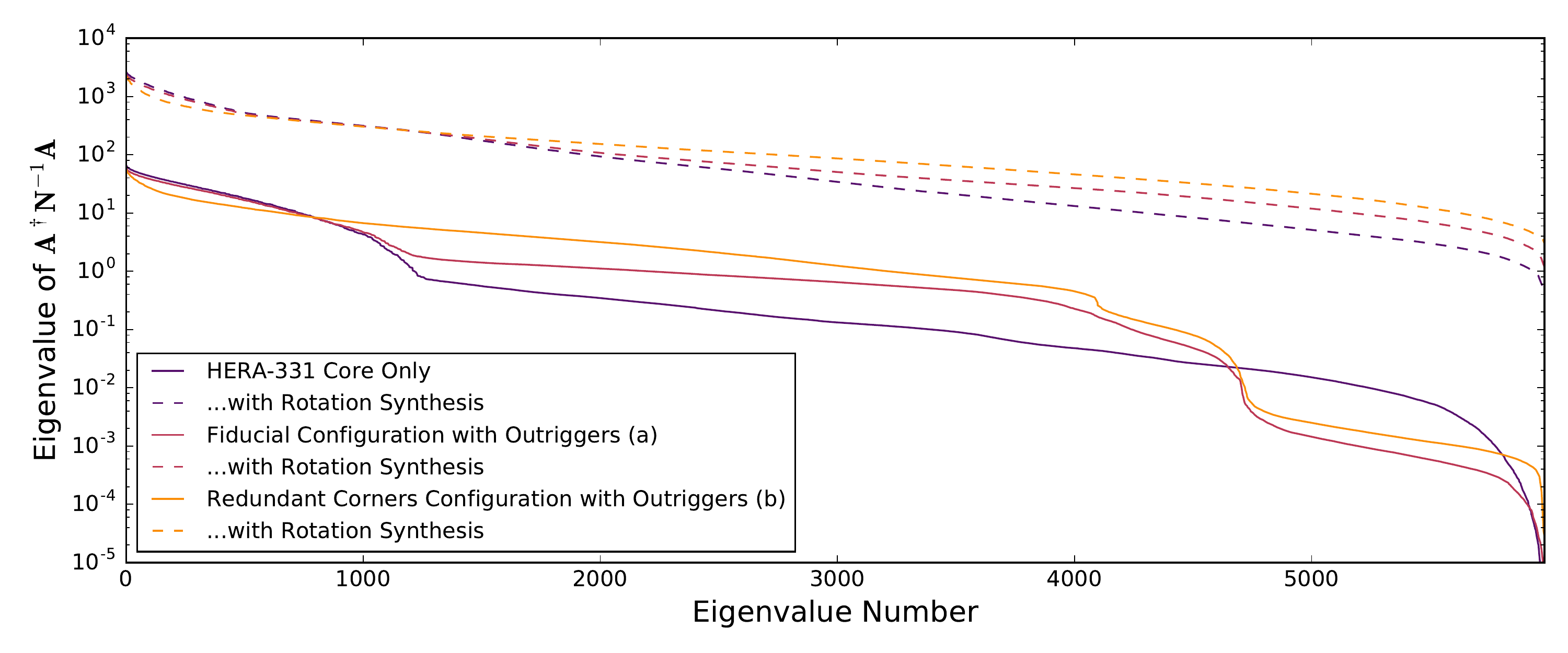}
	\vspace{-5pt}
	\caption{Here we examine the instantaneous and time-integrated high-resolution imaging capabilities of the arrays (with outriggers) shown in Figure \ref{fig:fullArrays}. Unlike in Figure \ref{fig:fullSkyEVs}, we calculate the PSF for only a $10^\circ$ facet, ignoring correlations with the rest of the sky, at HEALPix $N_\text{side} = 256$. We find that (b) and (c) concentrate their sensitivity on a relatively small number of modes while (c) spreads out the sensitivity. This effect is especially evident for instantaneous imaging, though still pretty modest. Since $\ANA$ is proportional to observing time, any shortcomings here can be erased by doubling the length of the observational campaign.}
	\label{fig:highResEVs}
\end{figure*}

We perform this analysis again on the full arrays (with outriggers) and obtain very similar results. In Figure \ref{fig:highResEVs} we assess the high-resolution mapmaking capability of each configuration by examining $\ANA$ for a $10^\circ$ facet around at the zenith at the midpoint of the observation at high resolution ($N_\text{side} = 256$).

We find that configurations (a) and (b), with their more compact cores, have more sensitivity to the best measured modes. By contrast, configuration (c) spreads its sensitivity out over a much larger number of independent modes. At very high eigenvalue number, we actually see configuration (a) performing well, due to its greater number of independently measured long baselines, although these modes are many orders of magnitude less sensitive than the best measured modes. Interestingly, even when rotation synthesis is included, the differences between the arrays are more persistent than they were in the widefield case. This is likely because the arrays differ more in their $uv$-coverage at long baselines than at short baselines, where the main hexagonal grid dominates.

\begin{figure*}[p] 
	\centering 
	\includegraphics[width=.85\textwidth]{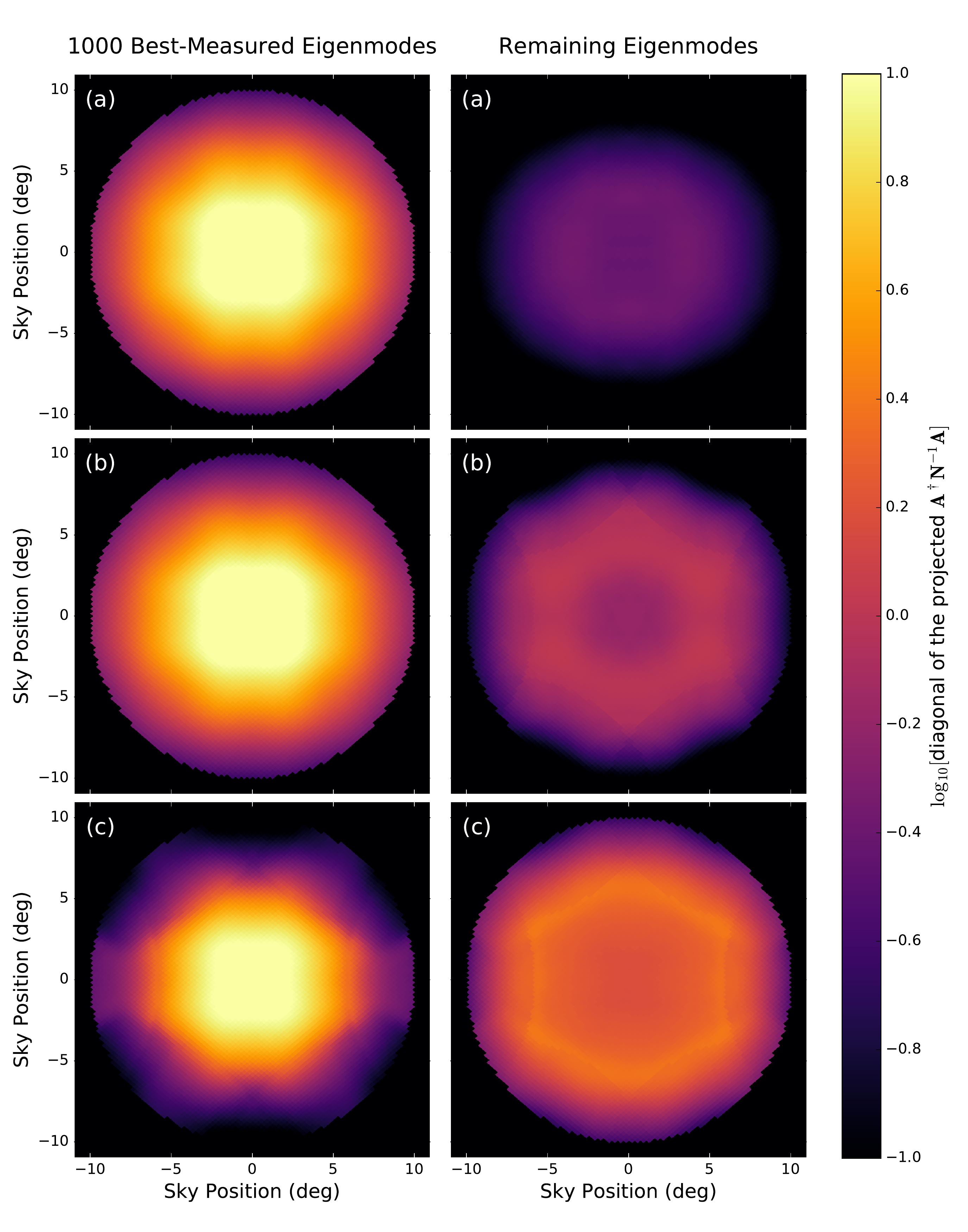}
	\caption{Just as in Figure \ref{fig:fullSkyProjected}, we can understand the significance of the main lobe, high-resolution snapshot eigenvalue spectrum of $\ANA$ in Figure \ref{fig:highResEVs} by looking that the diagonal elements of the matrix when it has been separated into two orthogonal subspaces. That separation---between the first 1000 eigenvalues and the rest---is again set by the crossing point where array configuration (c) begins out outperform the others. Once again, we see a cleaner separation between information in the center of the field and in the wide field. This supports the idea that configuration (c) sacrifices some sensitivity to the main lobe in favor of increased information about the sidelobes, which makes intuitive sense since the split-core configuration has observations more evenly distributed between on-grid and off-grid baselines (see Figures \ref{fig:redundancy} and \ref{fig:fullRedundancy}).}
	\label{fig:highResProjected}
\end{figure*}

Once again, by splitting the eigenspace of $\ANA$ for all three arrays at the eigenvalue crossing point of $N_\lambda \approx 1000$, we can get a sense of where the information comes from. Just as in Figure \ref{fig:fullSkyProjected}, we again see in Figure \ref{fig:highResProjected} that configuration (c) has a better separation of information between the center and the edge of the main lobe of the primary beam. Configuration (a), with almost no off-grid baselines, cannot separate information between zenith and the widefield very well at all.

Although the effects on mapmaking are modest, especially after Earth rotation synthesis, the separability of information between the main lobe of the primary beam and the sidelobes is interesting and potentially quite useful. In order to subtract foregrounds with sufficient fidelity to work within the wedge, we need precise subtraction of contaminants far from the main field of view \citep{PoberSidelobe, NithyaPitchfork}. The ability to instantaneously distinguish between foreground sidelobes and EoR signal is useful, especially when foreground positions and fluxes change due to the ionosphere \citep{HarishLeonIon1, HarishLeonIon2, InitialLOFAR1, AaronFirstEoXLimits}. Likewise, the cleaner separation between modes will likely improve traditional sky model-based calibration or hybrid calibration schemes that rely on a sky model for bandpass calibration.

\section{Redundant Calibratability} \label{sec:calibration}

Of course, good mapmaking is only one route to good calibration. If we cannot calibrate our instrument well, we risk introducing spectral structure into our visibilities and our maps---spectral structure that, when multiplied by our bright foregrounds, may contaminate the EoR window. Of the first generation of 21\,cm observatories, PAPER has so far produced the most constraining limits on the 21\,cm power spectrum, despite having the least collecting area. It owes that success to its relatively simple design, its tightly focused ``experimental'' approach that concentrates sensitivity on just a few baselines, and its use of redundant baseline calibration.

In this section, we begin by reviewing the mathematical formalism that underlies redundant baseline calibration in Section \ref{sec:calReview}. Then, in Section \ref{sec:calResults} we assess the redundant calibratability of the array configurations proposed and examined above. In particular, we show how even the off-grid antennas and outriggers can be redundantly calibrated.

\subsection{Redundant Calibration Review} \label{sec:calReview}

Fundamentally, the calibration problem for a radio interferometer can be expressed as\footnote{This leaves aside the question of ``direction-dependent calibration,'' which is essentially primary beam calibration. In this work, we assume our primary beams are the same. While it is possible to take antenna-dependent beams into account when making maps \citep{mapmaking}, it is not clear how much antenna-to-antenna variation \citep{AbrahamOrbcomm,NebenBeamformingErrors} affects redundant baseline calibration. It should also be pointed out that the sub-grid---and therefore sub-aperture---sampling enabled by all our configurations makes at least some kind of direction-dependent calibration possible using redundant information. Both questions are beyond the scope of this paper and are left for future work.}
\beq
V_{ij}^\text{measured} = g_i g_j^* V_{i-j}^\text{true} + n_{ij}. \label{eq:calibration}
\eeq
The visibility we measure, $V_{ij}^\text{measured}$, is related to the true visibility by the product of the complex gain $g_i$ of each antenna, plus some noise. We write the true visibility in shorthand as $V_{i-j}^\text{true}$ to express the idea that all that matters is the baseline, the separation between antennas $i$ and $j$, rather than the specific antennas involved. 

The key idea behind redundant baseline calibration is that the number of numbers we want to estimate, the gains and true visibilities, is much smaller than the number of measurements we actually make. We have an overdetermined system. If we can linearize Equation \ref{eq:calibration}, then we can write down a vector expression in the form of Equation \ref{eq:measurement} that relates all the visibilities measured to all the gains and true visibilities. Then we can use exactly the same statistical machinery to solve for both the gains and visibilities simultaneously.

\citet{redundant} propose two ways of doing that. The simpler way, called \texttt{logcal}, linearizes Equation \ref{eq:calibration} by taking the natural logarithm. If we express the complex gains as $g_j \equiv \exp[\eta_j + i\phi_j]$, then it follows that
\beq
\ln \left(V_{ij}^\text{measured}\right) = \eta_i + \eta_j + i\phi_i - i\phi_j + \ln\left( V_{i-j}^\text{true} \right) + n_{ij}'.
\eeq
The precise form and statistics of $n_{ij}'$ are discussed extensively in \citet{redundant}. This can be rewritten in the same form as Equation \ref{eq:measurement} by treating $\y$ as the set of measurements, $\x$ as a vector that contains both the gains and the true visibilities, and $\A$ as the matrix whose rows specify which pairs of antennas with which gains measure which visibility. In this form the real and imaginary parts of the gains and visibilities, produce two different sets of equations.

It turns out that \texttt{logcal}'s solutions are slightly biased. To rectify that, \citet{redundant} also developed \texttt{lincal}, a method of linearizing Equation \ref{eq:calibration} by Taylor expanding it and then solving for gains and visibilities iteratively. Both methods were implemented for \citet{MITEoR} and were integrated into the PAPER analysis pipeline \citep{PAPER64Limits}, with \texttt{logcal} providing \texttt{lincal} a good starting guess.\footnote{Both methods have been integrated into a single package and are available at \url{https://github.com/jeffzhen/omnical}}

In our case, we are interested in how the array configuration affects the errors on redundant calibration. For simplicity, we look at the expected gain errors predicted by $\texttt{logcal}$. If we form an unbiased estimator of the gains and true visibilities, then the gain/visibility covariance, $\boldsymbol\Sigma$, is given by
\beq
\boldsymbol\Sigma = \left[\ANA\right]^{-1}.
\eeq
In the case where all baselines measure visibilities with the same magnitude\footnote{This would be the case for a single dominant point source. At the frequencies and fields of view relevant to 21\,cm cosmology, that approximation would generally be a bad one. However, it is an illustrative limit to consider in our case when most visibilities are still the same order of magnitude.} and all antennas have the same noise level, \citet{redundant} show that $\boldsymbol\Sigma$ is reduced to  
\beq
\boldsymbol\Sigma \longrightarrow \frac{1}{(\text{S/N})^2} \left[\A^\dagger\A \right]^{-1},
\eeq
where S/N is the S/N of the sky signal (i.e.,~foregrounds), not just the 21\,cm signal. This leads us to a new metric for assessing array configurations by looking at the gain errors predicted by the diagonal of $\left[\A^\dagger\A \right]^{-1}$ It can be shown that the $\Delta \eta$s we predict are simply the gain errors up to third order in $\Delta \eta$:
\beq
\Delta|g_i| = \Delta \eta_i + \mathcal{O}(\Delta \eta^3).
\eeq

There remains one important complication related to the essential nature of what redundant calibration can and cannot do. As it turns out, $\A^\dagger\A$ is actually never invertible. There must always be at least one zero-eigenvalue because there is always an overall amplitude degeneracy in this system. We can increase the gains by some factor, decrease the true visibilities by that factor squared, and still get the same measurements. For the imaginary half of $\texttt{logcal}$, it turns out that there are always at least three zero-eigenvalues, one overall phase and two tip/tilt terms related to the orientation of the array.\footnote{In \texttt{lincal} or any other approach to redundant calibration, there will always be at least four zero-eigenvalues due to these degeneracies.} This limitation is discussed extensively in \citet{MITEoR}.

This is not to say that redundant baseline calibration is not helpful. It reduces the problem of calibrating a gain and a phase for every antenna at every frequency to calibrating just a few numbers per frequency for the entire array. The latter problem is less noisy, less sensitive to the accuracy of the sky model, and therefore far less error-prone. Fortunately, the gain errors we examine are insensitive to the overall amplitude, since they are normalized by the S/N. We merely need to take the pseudoinverse of $\A^\dagger\A$ to obtain $\boldsymbol\Sigma$ \citep{DillonFast}.

\subsection{Redundant Calibratability Results} \label{sec:calResults}

\begin{figure}[] 
	\centering 
	\includegraphics[width=.49\textwidth]{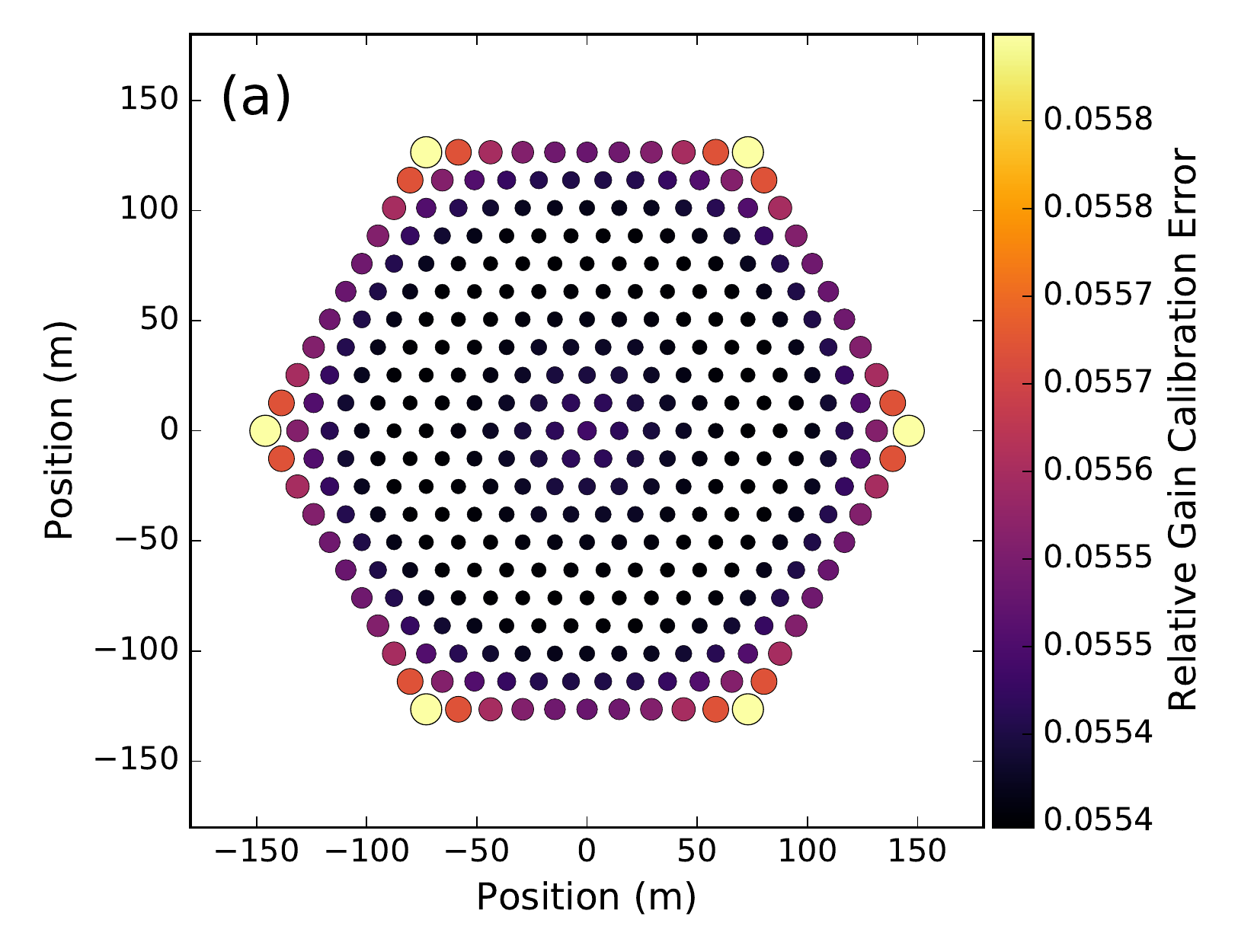}
	\caption{Here we show the expected error on the calibration (using both color and relative marker sizes) of antenna gains, relative to the S/N, for the core of configuration (a), the 331-element HERA hexagon. Using redundant baseline calibration, very precise gain solutions are attainable throughout. This calculation assumes that all baselines measure visibilities with the same amplitude, although in practice, amplitudes vary with baseline length, baseline orientation, and time. As in \citet{redundant}, we note that edge antennas are generally slightly less calibratable than interior antennas. Interestingly, the most calibratable antennas are off-center; this is because those antennas participate in a greater number of unique baseline types.}
	\label{fig:redCalCore}
\end{figure}

Before we look at the three array configurations considered above, let us try to develop some intuition for intuition for this formalism. In Figure \ref{fig:redCalCore}, we plot the gain errors expected from redundant calibration of a 331-element solid hexagonal HERA core, the core of configuration (a).
Plotted are the gain errors we would expect if each visibility had the same amplitude and were measured with a foreground S/N of 1. The errors depend inverse-linearly on the S/N. 

\begin{figure*}[t] 
	\centering 
	\begin{tabular}{ccc}
  \includegraphics[width=.49\textwidth]{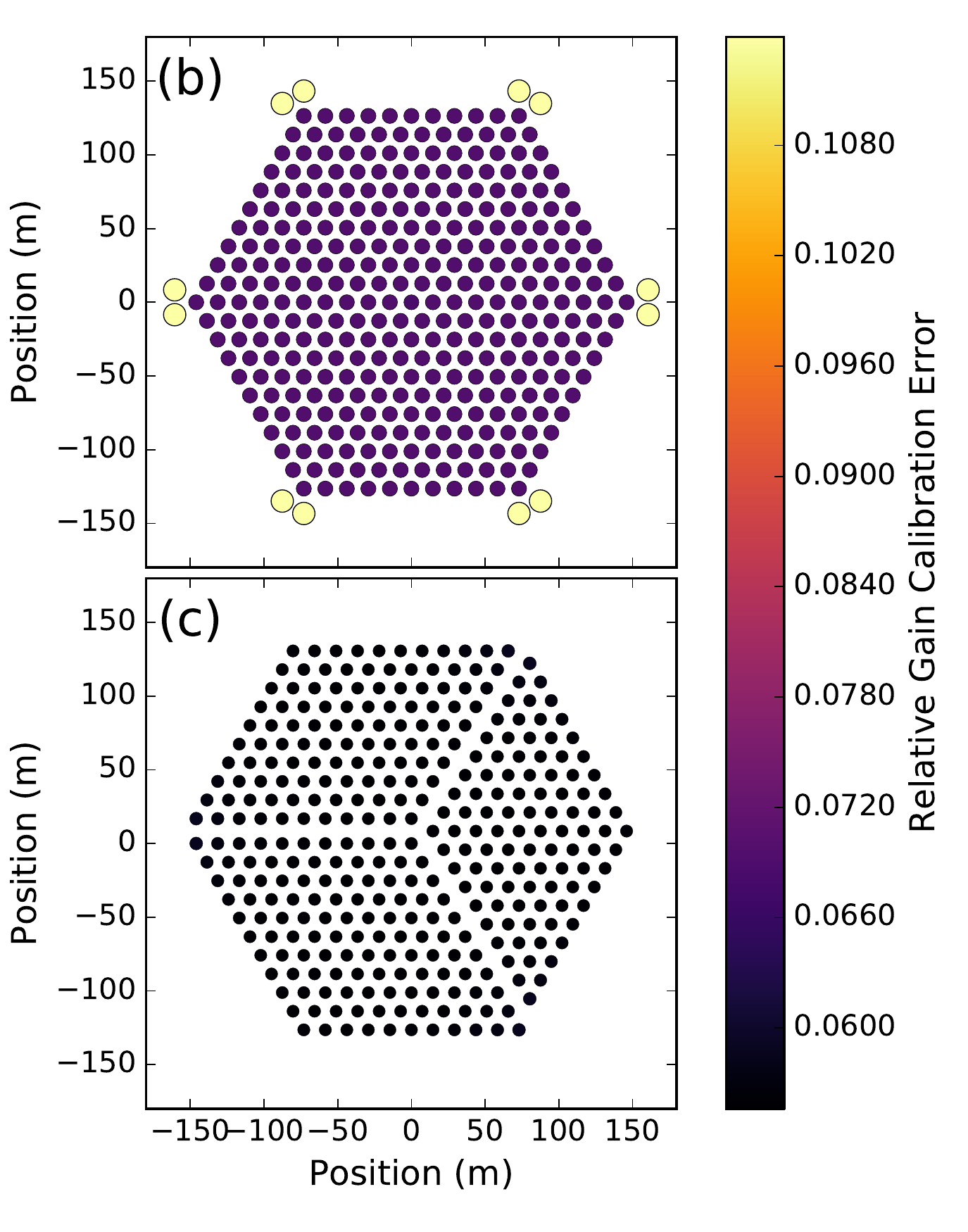} & 
\hspace{.02\textwidth} &
  \includegraphics[width=.49\textwidth]{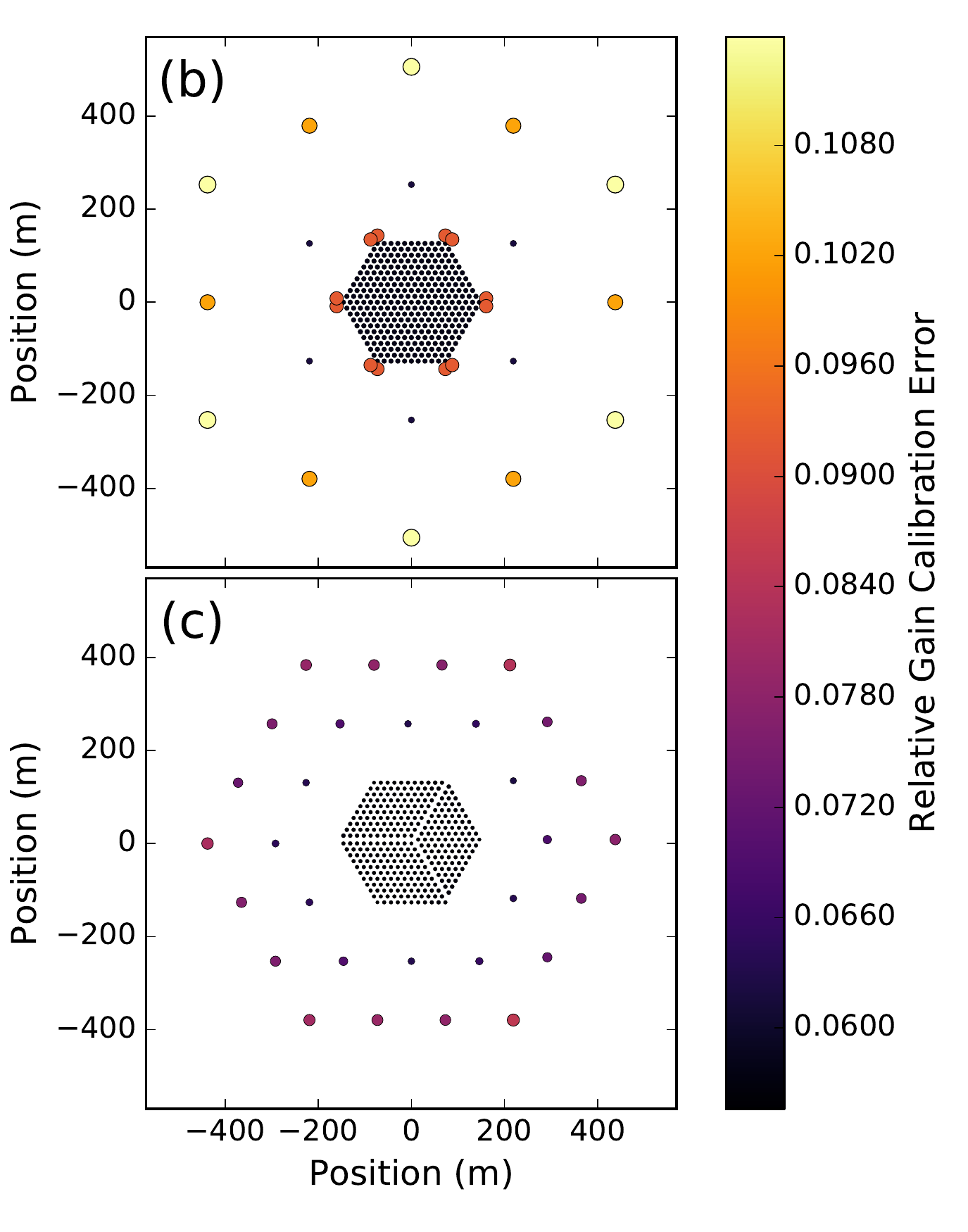}
        \end{tabular}
\vspace{-15pt}
	\caption{Expected gain errors relative to S/N (shown with both color and relative marker size) for configurations (b) and (c). On the left we show the relative gain errors of the cores alone, on the right we show all the errors when calibrating the full arrays using redundant calibration. When only calibrating the cores, configuration (b) shows increased gain errors on the whole array by about 20\%. The dithered corner antennas, in particular, have roughly double the error of the rest of the array, meaning that all off-grid baselines will be effectively noisier by a factor of 2 (in the visibilities). Configuration (c) has almost the same errors in the core as an unmodified hexagonal core. When outriggers are added, the differences between the cores are reduced to only about 5\%. However, the furthest outriggers have considerably larger error in configuration (b) than configuration (c). Note that configuration (a) is excluded here because it is not redundantly calibratable with its outriggers.}
	\vspace{5pt}
	\label{fig:redCalAll}
\end{figure*}

For a highly redundant configuration like the solid HERA core, the antenna-to-antenna variation in gain errors is very small. In these cases, the relative gain errors are all roughly just $(N_\text{antennas})^{-1/2}$ \citep{redundant}. However, we can see some illustrative structure in Figure \ref{fig:redCalCore}. The edge and especially corner antennas are are slightly more difficult to calibrate---they are involved in fewer highly redundant baselines than interior antennas. Likewise, antennas in the very center are involved in fewer kinds of baselines and thus also see slightly higher error. Both the redundancy of the baselines an antenna is involved in and the centrality of those baselines to the larger connected graph of redundant baselines affect the calibratability of each antenna. Regardless, for a 331-element core these variations are small in comparison to what we see for more complex array designs; the key comparison we wish to make is how well we can redundantly calibrate our full arrays relative to the a solid core.

After calculating the \texttt{logcal} $\A^\dagger\A$ for each of the three arrays we have considered above, configuration (a) sticks out right away. The imaginary version $\A^\dagger\A$ should have three zero-eigenvalue; it has five. Two extra phase factors are not redundantly calibratable because the outrigger baselines do not overlap as they do in configuration (b). While the core of configuration (a) can be redundantly calibrated, its outriggers cannot. We therefore cannot predict the errors from redundant calibration for this array. 

Configurations (b) and (c), however, were both designed to be redundantly calibratable.\footnote{Even in the event that a single antenna breaks, they can still be redundantly calibrated.} They have the correct number of zero-eigenvalues, meaning that taking the pseudoinverse of $\A^\dagger\A$ is appropriate. In Figure \ref{fig:redCalAll} we show the expected errors on antenna gains when calibrating the cores redundantly and when calibrating the whole array redundantly.

When calibrating just the cores, the addition of the 12 off-grid corner antennas in configuration (b) leads to higher gain errors not just in the off-grid antennas but also, surprisingly, in the hexagonal core by about 20\%. By contrast, configuration (c) sees raised errors generally by less than 1\% compared to a solid HERA core. This is because the core of configuration (b) probes many more baselines than a simple HERA core, but most of them are sparsely sampled. A large number of baselines are only weakly connected to the rest, leading to higher errors throughout. Of course, if one wishes to give up on redundant calibration of the off-grid antennas, one can achieve the same errors as in Figure \ref{fig:redCalCore}. However, then one must resort to another strategy (i.e. one that requires a sky model) to calibrate the remaining antennas (which affect all the off-grid visibilities).

Those effects are mitigated somewhat by the addition of redundant outriggers that decrease the discrepancy between the array cores to a 5\% difference in calibration errors. On the other hand, the most distant outriggers in configuration (b) have approximately 50\% larger gain errors than the most furthest-flung outriggers in configuration (c). Simply put, the evenness of the distribution of redundancy across the baselines probed, and thus the connectedness of antennas through that network of redundant baselines, makes configuration (c) more redundantly calibratable.


\section{Summary of Array Design Lessons} \label{sec:summary}

In this work, we examined how minor modifications to interferometric array configurations with dense cores optimized for 21\,cm cosmology can both produce useful images and be redundantly calibrated, even though they are designed for foreground avoidance and maximal sensitivity in the EoR window. Based on \citet{mapmaking} and \citet{redundant}, we quantified these effects by examining the information content or, relatedly, the expected error. These are two new ways to think about array configurations and we expect them to be broadly useful for designing future 21\,cm arrays that rely on dense cores and redundant calibration. 

The differences between the arrays we examined for mapmaking capability and redundant calibratability were relatively small, although we can attribute this to the restriction that we maintain a high-sensitivity, dense core optimized for foreground avoidance---currently the most robust and promising approach for accessing the high-redshift 21\,cm signal. Despite that, we can extrapolate a number of relevant lessons for future telescopes:
\begin{enumerate}
\item As expected, adding ``off-grid'' baselines improves instantaneous widefield imaging and the separation between modes in the main lobe of the primary beam and those further from the zenith.
\item A redundantly calibratable array need not necessarily be bad for imaging. A redundantly calibratable array can still have dithered, off-grid baselines. It can still have distant outriggers. And even though redundant calibration relies on measuring only a relatively modest number of unique visibilities at any given time, we can get back sensitivity to a large number of distinct modes with Earth rotation synthesis.
\item Redundant calibratability does not simply depend on the number of antennas and the number of unique visibilities. The ``connectedness'' of the relationship between gains and visibilities---which well-measured visibilities are most useful for constraining which gains and vice versa---is important as well. This is a reason to prefer configurations like (c) over simply adding a few off-grid antennas.
\item The linear estimator formalism that underlies both calculations in this paper is a powerful tool for assessing the expected performance of an instrument during calibration and data reduction.
\end{enumerate}

In time, it would be useful to examine how these effects propagate to the scientific results we expect from instruments like HERA: 21\,cm power spectra and constraints on reionization. While \citet{GreigSKADesign} was able to perform this analysis for widely differing SKA configurations, they relied on approximate sensitivity calculations that ignore correlations between $uv$-cells---precisely the effect that matters when we start examining off-grid and therefore sub-aperture baselines. To our knowledge, no power spectrum analysis has been demonstrated that takes into account full noise correlations and the frequency dependence of both the primary beam and the PSF. A rigorous and more quantitative examination of array configurations and their effect on astrophysical and cosmological constraints is left to future work. However, our new techniques show how to make small sacrifices in sensitivity in order to obtain sizable improvements in the control of systematics. With better instrument simulation and power spectrum estimation, one can assess the metrics we have developed here to understand their relative effects on 21\,cm cosmology.

\section*{Acknowledgements}
The authors gratefully acknowledge the contributions of a number of colleagues and collaborators. Adrian Liu was extremely helpful in giving feedback and general advice about the direction of this work. Ue-Li Pen gave us ideas that eventually led to configuration (c). Jeff Zheng helped us use his \texttt{Omnical} package and double-checked the redundant calibratability of configurations (b) and (c). Likewise, Jonathan Pober helped us apply his \texttt{21cmSense} package. Miguel Morales and Bryna Hazelton gave us useful feedback on how to examine PSFs and sidelobe suppression. David DeBoer gave us feedback on the feasibility of these array configuration modifications (and more radical ones not considered in this work) and, with  James Aguirre, supplied us with HERA primary beam simulations. Finally, we would like to thank Max Tegmark for originally sparking an interest in the question of array configuration optimization, starting with \citet{FFTT2}, and continuing with discussions for years afterwards.

This research was conducted
as part of the University of California Cosmic Dawn
Initiative. The authors acknowledge support from the University
of California Office of the President Multicampus
Research Programs and Initiatives through award MR-15-
328388. The authors also acknowledge support form the National Science Foundation through CAREER award No.\ 1352519, AST grant No.\ 1129258, and AST grant No.\ 1440343.

\bibliography{ArrayConfiguration}{}
\end{document}